\documentclass[aps,prl,twocolumn, groupedaddress]{revtex4-1}
\usepackage[pdftex]{graphicx}
\usepackage{subfigure}
\usepackage{sidecap}

\usepackage{graphicx}
\usepackage[intlimits]{amsmath}
\usepackage{amssymb}
\usepackage{multirow} 

\usepackage{siunitx} 
\def\percent{\%}
\def\decibel{dB}

\begin{document}


\title{Audio-band frequency-dependent squeezing}



\def\vup{\vspace{-2pt}}

\author{Eric Oelker}
\author{Tomoki Isogai}
\author{John Miller}
\author{Maggie Tse}
\author{Lisa Barsotti}
\author{Nergis Mavalvala}
\author{Matthew Evans} \email{mevans@ligo.mit.edu}
\affiliation{Massachusetts Institute of Technology, Cambridge, MA 02139, USA \vup}

\date{\today}


\newcommand{\Fref}[1]{Fig.~\ref{fig:#1}}
\newcommand{\Tref}[1]{Table.~\ref{tab:#1}}

\def\Pthresh{\bar{P}}
\def\Parm{P_{\rm arm}}

\def \gw {gravitational-wave}
\def \fin {\mathcal{F}}


\begin{abstract}
  Quantum vacuum fluctuations impose strict limits on precision
  displacement measurements, those of interferometric
  gravitational-wave detectors among them.  Introducing squeezed
  states into an interferometer's readout port can improve the
  sensitivity of the instrument, leading to richer astrophysical
  observations.  However, optomechanical interactions dictate that the
  vacuum's squeezed quadrature must rotate by 90 degrees around
  \SI{50}{Hz}.  Here we use a 2-m-long, high-finesse optical resonator
  to produce frequency-dependent rotation around \SI{1.2}{kHz}.  This
  demonstration of audio-band frequency-dependent squeezing uses
  technology and methods that are scalable to the required rotation
  frequency, heralding application of the technique in future
  gravitational-wave detectors.
\end{abstract}


\maketitle


\section{Introduction}
\label{sec:intro}

Quantum vacuum fluctuations permeate the entirety of space.
Ordinarily benign, these jittering fields impose the strictest limit
on the precision of microscopic measurements.  In particular, quantum
noise limits the performance of interferometric gravitational-wave
detectors as they attempt to make the first observations of ripples in
the very fabric of space-time~\cite{Fritschel2015, Aso13a, Deg12a}.
By squeezing the vacuum fluctuations entering an interferometer's
readout port, the sensitivity of the instrument can be significantly
improved~\cite{Teich1989, Cav81a, Bar13a, LIG11b}, leading to richer
astrophysical observations.  However, optomechanical interactions
dictate that the vacuum's squeezed quadrature must rotate by 90
degrees around \SI{50}{Hz}, equivalent to storing the entangled
photons for \SI{3}{ms}~\cite{Aspelmeyer2014,Kimble01, Kwee14a}.  Here
we store spectral components of a squeezed state for
\SI{128}{\micro\second} in a 2-m-long, high-finesse optical resonator
to produce frequency-dependent rotation at \SI{1.2}{kHz}.  This first
demonstration of frequency-dependent squeezing in the audio-band uses
technology and methods that are scalable to the required rotation
frequency, heralding application of the technique in all future
gravitational-wave detectors~\cite{Miller2015a}.

Just as the ground state of the quantum harmonic oscillator has non-zero
energy and an associated uncertainty principle, so too does that of
the electromagnetic field. In the latter case, the ground state energy
gives rise to so-called quantum vacuum fluctuations in the field and
the accompanying uncertainty principle relates the variances in its two
orthogonal quadrature phases.

Although seemingly insignificant, quantum vacuum fluctuations impose
the principal limit on the sensitivity of present-day \gw\
interferometers. Both low-frequency radiation pressure noise and
high-frequency shot noise arise due to the vacuum fluctuations which
enter an interferometer's readout port~\cite{Cav81a}.

The vacuum state naturally present in all modes of the electromagnetic
field possesses equal uncertainty in each of its two quadratures.
However, it is possible to redistribute the uncertainty, in accordance
with the Heisenberg Uncertainty Principle, to produce a
\emph{squeezed} state, with reduced variance in one quadrature at the
expense of increased variance in the orthogonal quadrature (see
\Fref{apparatus}). Common techniques used include parametric
down-conversion~\cite{Furst2011}, four-wave
mixing~\cite{Shelby1986,Lambrecht1996,Ourjoumtsev2011}, the Kerr
effect~\cite{Nishizawa1994,Fox1995} and nonlinearities in
optomechanical systems~\cite{Brooks2012,Safavi2013,Purdy2013}. The
most advanced sources are currently optical parametric oscillators
(OPOs)~\cite{Vahlbruch2008,Chua2011}. The bandwidth of these devices
is $\gtrsim$\SI{10}{MHz}, so that the redistribution of noise occurs
essentially uniformly over the range of frequencies of interest for
\gw\ detectors.

In 1981 Caves proposed the injection of squeezed vacuum, in place of
coherent fluctuations, in order to reduce high-frequency shot noise in
\gw\ interferometers~\cite{Cav81a}. This idea came to fruition only in
the current decade with successful demonstrations at the GEO600 and
LIGO Hanford detectors~\cite{LIG11b,Bar13a}, proving that
squeezed-light injection could be performed without degrading the
exquisite low-frequency sensitivity of these instruments.

Today, squeezed vacuum sources offer squeezing magnitudes of more than
\SI{10}{\decibel} (corresponding to approximately a 3-fold reduction in noise
amplitude) and, critically, maintain this performance down to
\SI{10}{Hz}~\cite{Mehmet2011, Stefszky2012}. Additionally, a squeezed vacuum
source is permanently installed at GEO600, enabling investigations of
long-term robustness~\cite{Grote2013} and the evaluation of different
control schemes~\cite{Dooley2015}.

Building on this foundation, it is possible to utilize this proven
technique to mitigate quantum noise in current kilometer-scale \gw\
interferometers. However, a simple frequency-independent squeezed
vacuum source is not sufficient for the present generation of
detectors~\cite{Eva13a}. To realize broadband noise reduction one must
rotate the squeezed quadrature as a function of frequency in order to
counter the rotation effected by the optomechanical coupling between
the interferometer's \SI{40}{kg} mirrors and the nearly \SI{1}{MW} of
circulating laser light~\cite{BnC}.

\section{Production of Frequency Dependent Squeezing}
\label{sec:production}

The appropriate rotation can be achieved by reflecting a standard
frequency-independent squeezed vacuum state off a low-loss optical
resonator known as a filter
cavity~\cite{Kimble01,Har03a,Khalili2007,Kwee14a}. As with other
resonances, that of the filter cavity is dispersive. Spectral
components of the squeezed vacuum that lie within the linewidth of the
cavity experience a change in their phase upon reflection; those
outside the linewidth do not. By operating the filter cavity in a
detuned configuration, differential phase can be imparted upon the
upper and lower squeezed vacuum sidebands, leading to frequency
dependent quadrature rotation.

\begin{figure*}[htbp!]
    \includegraphics[width=\textwidth]{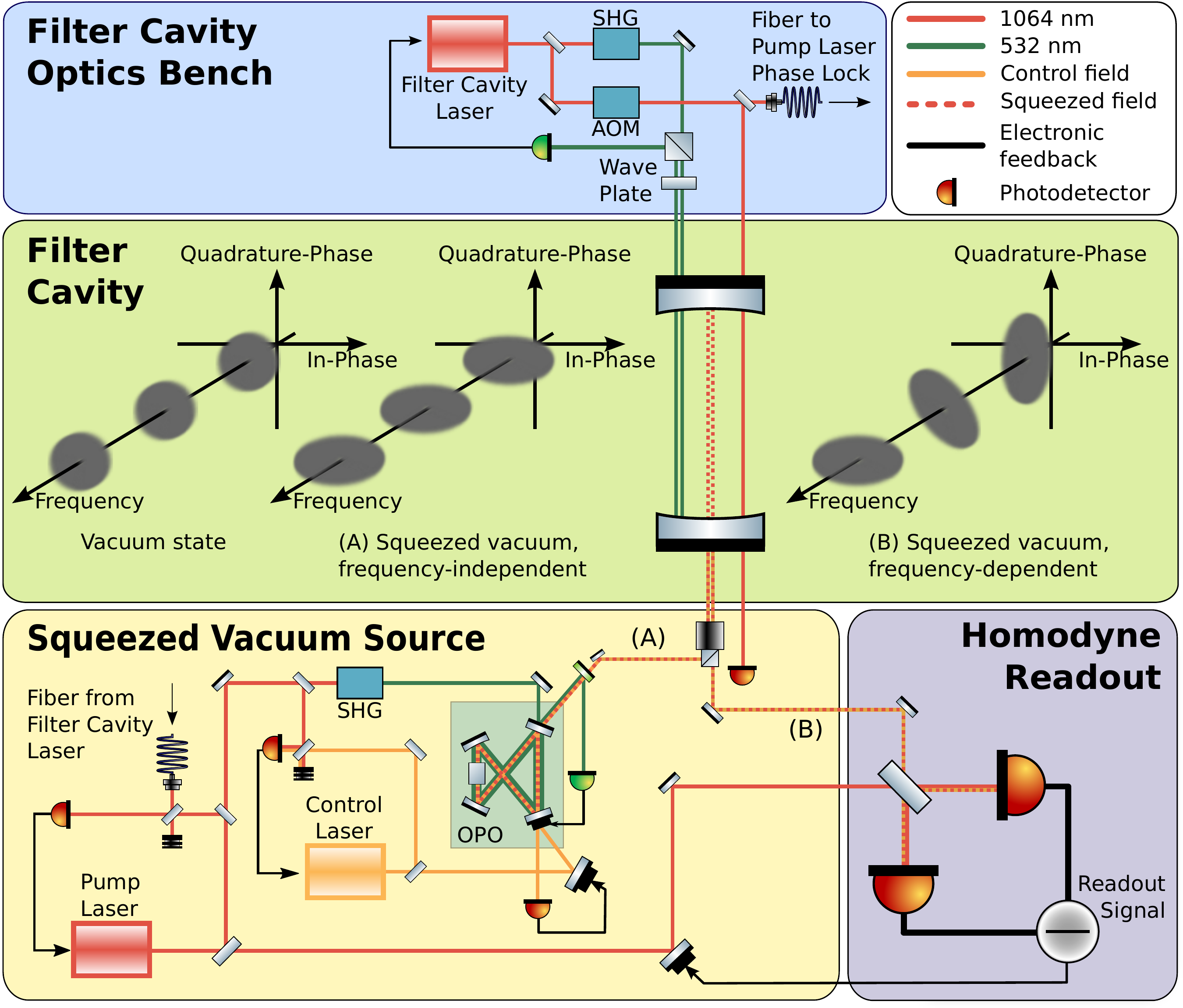}
    \caption{Audio-band frequency-dependent squeezed vacuum
      source. Frequency-independent squeezed vacuum is produced using
      a dually-resonant sub-threshold OPO operated in a traveling-wave
      configuration.  The OPO is pumped with light provided by a
      Second Harmonic Generator (SHG). The generated squeezed state is
      subsequently injected into a dichroic (1064/532~nm) filter
      cavity along path (A) where it undergoes frequency-dependent
      quadrature rotation. A Faraday isolator redirects the returning
      squeezed field along path (B) towards a homodyne readout system
      where frequency-dependent squeezing is measured. The control
      laser is phase locked to the pump field with a detuning of 29.5
      MHz and is injected through the rear of the OPO cavity. This
      field co-propagates with the squeezed field to the homodyne
      readout and is used to lock the local oscillator phase using the
      coherent control technique. The detuning of the filter cavity is
      set using an acousto-optic modulator (AOM).}
    \label{fig:apparatus}
\end{figure*}


The cavity detuning effectively determines the relative delay between
the photons in each entangled pair. While we implemented this
adjustable delay using an auxiliary laser source (see
\Fref{apparatus}), other potential means of adjusting the delay
involve tuning of the control sideband modulation~\cite{Yam2015} or
employing a variable finesse cavity~\cite{Barsotti2006, Hild2009}.

The frequency range over which rotation takes place is set by the
filter cavity storage time $\tau_\text{storage}$.
\begin{equation}
  \tau_\text{storage} = \frac{1}{\gamma_\text{fc}},
\end{equation}
where $\gamma_\text{fc} = \pi c/(2L_\text{fc}\fin)$ is the
half-width-half-maximum-power linewidth (in radians per unit time) of
a cavity with length $L_\text{fc}$ and finesse $\fin$, $c$ being the
speed of light.

To implement frequency dependent squeezing in Advanced
LIGO~\cite{Fritschel2015} a \SI{3}{ms} storage time is
required; comparable to the longest ever recorded~\cite{DellaValle14}.
As optical losses severely limit the finesse and storage time
achievable for a given cavity length~\cite{Isogai13}, experimental
realization of such cavities is extremely challenging, with the only
prior demonstration of quadrature rotation in this way having targeted
MHz frequencies~\cite{Chelkowski2005}. Nevertheless, filter
cavities represent the best prospect for developing an audio-band
frequency-dependent squeezed vacuum source in the near future, with
other techniques restricted by thermal noise~\cite{Qin2014}, the level
of squeezing produced at low frequencies~\cite{Corzo2013} or
loss-induced decoherence~\cite{Horrom2013}.

In this work we describe the first instance of frequency-dependent
squeezing at audio frequencies. By injecting light from a squeezed
vacuum source into a filter cavity and measuring the spectrum of noise
in the reflected field as a function of quadrature phase, we
demonstrate $90^\circ$ quadrature rotation of a squeezed vacuum state at
\SI{1.2}{kHz}. This is the first demonstration of the desired $90^\circ$ rotation required
 for Advanced LIGO and other similar \gw\ detectors, as well as the first demonstration of quadrature
  rotation of a squeezed \emph{vacuum} state.\footnote{As opposed to a ``bright'' squeezed state,
   which has non-zero mean field amplitude.  Vacuum squeezed states are necessary to avoid technical
  noises in sub-MHz measurements, as required by \gw\ detectors~\cite{Schnabel2004}.}
This result establishes frequency-dependent squeezing as a viable technique
 for improving the sensitivity of \gw\ interferometers.

\begin{table}[tbhp!]
\centering
\caption{Parameters of our frequency-dependent squeezed vacuum source. Entries marked by an asterisk were determined most accurately through fitting to recorded data. In all cases fitting produced values in keeping with independent measurements and their uncertainties.}
\label{tab:param}
\begin{tabular}{  l  c  }
  \hline
  Parameter                          & Value\\
  \hline                                                   
  \hline
  Filter cavity length               & \SI{1.938408 \pm 0.000006}{m}\\
  Filter cavity storage time         & \SI{127.5 \pm 2.5}{\micro\second}\\
  Filter cavity decoherence time     & \SI{1.8 \pm 0.4}{ms}\\
  OPO nonlinear gain$^*$             & \SI{12.7 \pm 0.4}{}      \\
  OPO escape efficiency              & \SI{95.9 \pm 1}{\percent}                   \\
  Propagation loss$^*$               & \SI{11.9 \pm 0.9}{\percent}    \\
  Homodyne visibility                & \SI{96.6 \pm 1}{\percent}                   \\
  Photodiode quantum efficiency      & \SI{93 \pm 1}{\percent}                     \\
  Filter cavity round-trip loss$^*$       & \SI{7.0 \pm 1.6}{ppm}                   \\
  Freq. indep. phase noise (RMS)$^*$ & \SI{31 \pm 7}{mrad}   \\
 Filter cavity length noise (RMS)$^*$     & $(7.8\pm0.2)\times10^{-13}$ m     \\
Filter cavity-squeezed vacuum        & \multirow{2}{*}{\SI{97 \pm 2}{\percent}}                     \\
 mode coupling                       &\\
  \hline
\end{tabular}
\end{table}

Our experimental setup is depicted in \Fref{apparatus}. The apparatus
consists of a broadband squeezed vacuum source, a detuned filter
cavity producing the desired frequency-dependent rotation of the
squeezed quadrature, ancillary systems which set the detuning of the
filter cavity, and a balanced homodyne detection system for measuring
the squeezed state.  Key parameters of the system are listed in
\Tref{param}.

The squeezed vacuum source is built around a traveling-wave OPO
cavity~\cite{Chua2011,Stefszky2012} resonant for both the \SI{532}{nm}
pump light and the \SI{1064}{nm} squeezed vacuum field it generates.
Our OPO outputs \SI{11.8\pm0.5}{\decibel} of squeezing via parametric
down-conversion in a non-linear periodically-poled KTP crystal.  After
leaving the OPO, the squeezed field is reflected off the filter
cavity, inducing rotation of the squeezed quadrature for spectral
components that lie within the cavity linewidth.

The filter cavity is a symmetric, near-concentric, 2 m long
Fabry-Perot cavity.  It has a storage time of \SI{128}{\micro\second}
and a finesse of $\sim$30000 for 1064 nm light.  The inferred cavity
round-trip loss, excluding input coupler transmissivity, is
$L_\text{rt}=7$~ppm~\cite{Isogai13}, corresponding to a decoherence
time, defined by
\begin{equation}
  \label{eq:decoherence}
  \tau_\text{decoherence}=\frac{-2L_\text{fc}}{c\ln(1-L_\text{rt})},
\end{equation}
of \SI{1.8}{ms}. The cavity also features a low-finesse ($\sim$150)
532 nm resonance which is used to stabilize the detuning of the
squeezed field relative to the filter cavity.

Finally, balanced homodyne detection \cite{Stefszky2012, Fritschel:14}
is used to measure the squeezed state after reflection from the filter
cavity.  The output of the homodyne detector is used to fix the
quadrature of the squeezed state relative to the local oscillator
field using the coherent control technique~\cite{Chelkowski07}.

\begin{figure*}[t!]  \centering
   \includegraphics[width=0.95\textwidth]{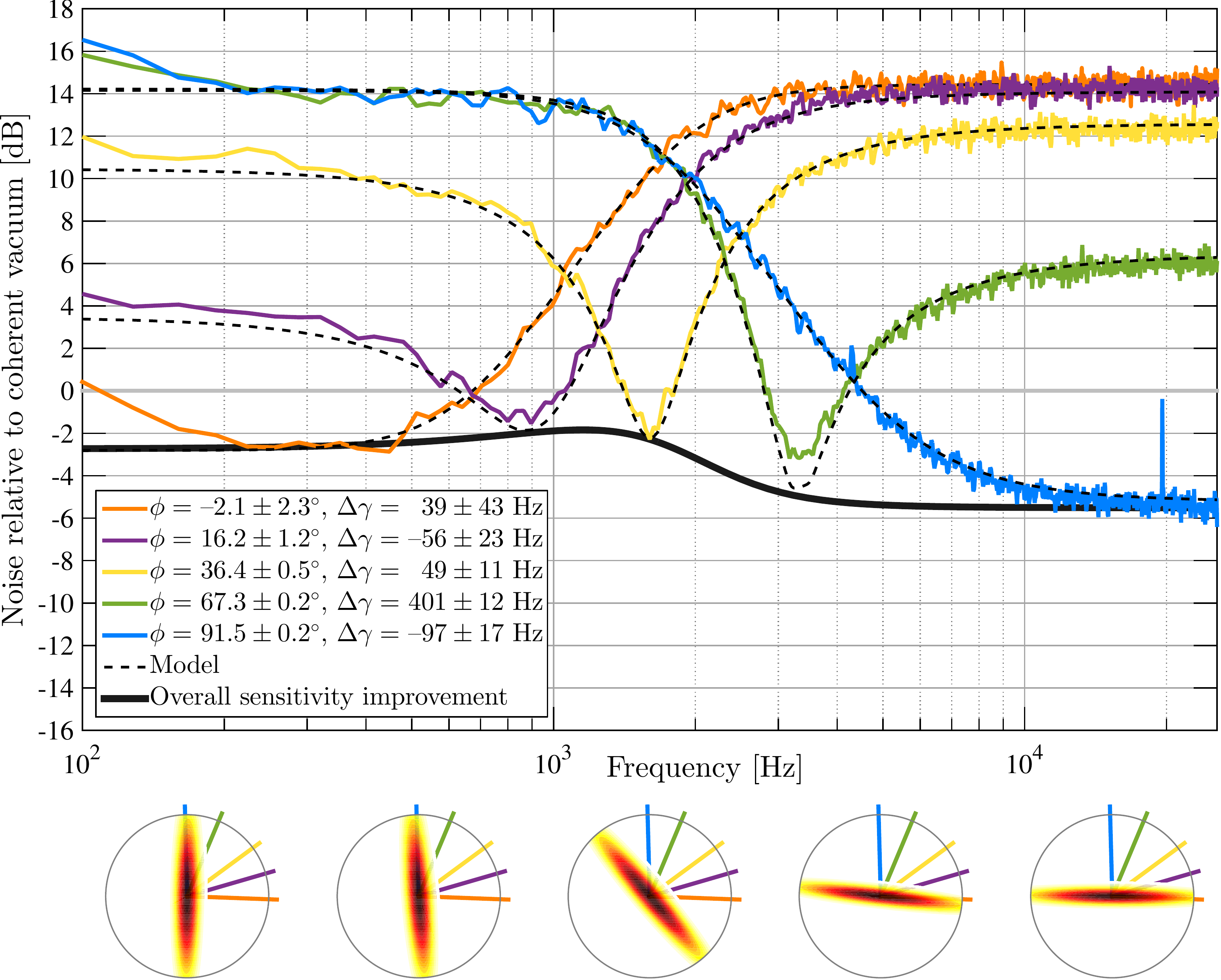}
   \caption{Demonstration of frequency-dependent squeezing. Measured
     noise relative to that due to the naturally occurring vacuum
     state (shot noise) as a function of sideband frequency and
     readout quadrature $\phi$. The difference between the
     experimental filter cavity detuning (offset from resonance) and
     the desired value of \SI{1248}{Hz} is denoted by
     $\Delta\gamma$. Dashed curves represent the output of the model
     described in~\cite{Kwee14a} using the parameters given in the
     legend and \Tref{param}. The deviation of the recorded data from
     prediction below $\sim$300Hz is due to unmodeled environmental
     disturbances rather than a fundamental limitation of the
     technique. Data in this band were excluded from our analyses.
     The solid black curve provides an estimate of the overall
     improvement achievable if this frequency-dependent squeezed
     vacuum source were applied to the appropriate
     interferometer. Ellipses illustrate the modeled squeezing
     magnitude and angle (Wigner function) at the frequency at which
     they are located. The measured noise is given by the projection
     of the ellipse onto the appropriately coloured readout vector.}
    \label{fig:results}
\end{figure*}

Measured quantum noise spectra are presented in \Fref{results}. The
data are normalized with respect to the value detected with unsqueezed
vacuum fluctuations such that the reported values describe the
deviation from shot noise due to the addition of squeezing.

Rotation of the squeezed quadrature occurs near
\SI{1}{kHz}.  Squeezing levels of \SI{5.4\pm0.3}{\decibel} and
\SI{2.6\pm0.1}{\decibel} are observed at high and low frequency,
respectively.  Weaker squeezing at low frequencies is due to the
spectral selectivity and internal losses of the filter cavity, which
result in some decoherence of the squeezed state.  To achieve the
desired quadrature rotation, the central frequency of the squeezed
vacuum field is held close to filter cavity resonance; low-frequency
squeezing sidebands thus interact with the filter cavity whereas
high-frequency sidebands are reflected and incur very little loss.

A quantum noise model that includes realistic, frequency-dependent
decoherence and degradation mechanisms was used to evaluate our
results~\cite{Kwee14a}.  While all aspects of our system were
meticulously characterized, certain parameters were most accurately
quantified through fitting this model to the measured data (see
Table~\ref{tab:param}).  With the exception of filter cavity detuning
and readout quadrature angle, which are different for each of the five
measurements reported, a single value for each system parameter was
determined using all available data sets.  In all cases, fitted values
were consistent with direct measurements, given their uncertainties.
Furthermore, the close agreement between our measured data and the
quantum noise model, presented in \Fref{results}, indicates that all
significant sources of squeezed state decoherence and degradation are
well modeled.

\subsection{Control of Filter Cavity Detuning}
\label{sec:technical}

The output of a \SI{500}{mW}, \SI{1064}{nm} Nd:YAG
solid state laser (filter cavity laser, \Fref{apparatus}) was split
into two portions: the first was frequency doubled and used to lock
the laser to the filter cavity with a bandwidth of
\SI{30}{kHz} (the filter cavity mirrors are dichroic, with a
finesse of $\sim$150 at \SI{532}{nm}); the second was
double-passed through an acousto-optic modulator (AOM) and used to
phase lock the pump laser to the filter cavity laser. The result of
this setup is that the frequency offset of the pump laser, and
therefore the squeezed vacuum, from resonance in the filter cavity
was 
stabilised and could be controlled by varying the AOM's drive
frequency. In order to produce the desired $90^\circ$ quadrature
rotation, the offset was set to the filter cavity
half-width-half-maximum-power frequency.

\subsection{Impact of Technical Noise}

Optical loss, mode mismatch, and squeezed quadrature fluctuations (or
phase noise) cause decoherence and a reduction in measurable
squeezing~\cite{Dwyer13,Kwee14a}.  For instance, an ideal squeezed
vacuum source with our operating parameters should produce
\SI{15.6}{\decibel} of squeezing, yet, as expected when the
deterioration due to the above listed effects is taken into account,
the level we measured was below \SI{6}{\decibel}.

Each source of loss leads to decoherence of the entangled photons
which make up a squeezed vacuum state.  Losses outside the filter
cavity affect all frequencies equally and arise due to imperfect
optics (propagation loss, OPO escape efficiency), non-unity photodiode
quantum efficiency, and imperfect mode overlap between signal and
local oscillator beams at the homodyne detector (visibility). One may
combine all of these factors in to a single \emph{detection loss},
which was \SI{29}{\percent} in our system.  Detection loss can be
reduced, though at significant cost, through use of improved
polarization optics~\cite{Skeldon2001}, superior photodetectors and
active mode matching~\cite{Mueller00}.

The treatment of filter cavity losses is more complicated due to their
frequency dependence~\cite{Kwee14a}. As an indication, the total loss
of our cavity was approximately \SI{16}{\percent} on
resonance. Advances in this area are limited by the combination of
currently available cavity optics and the necessity of using long
cavities, and thus large mm-scale beams, to achieve the required
storage times~\cite{Isogai13}.

The mode coupling between the squeezed field and the filter cavity
must also be considered. This effect is more subtle than a simple loss
since that portion of the squeezed vacuum not matched to the filter
cavity is not rotated in the desired manner and yet still arrives at
the homodyne detector. This ``dephasing'' effect corrupts the squeezed
quadrature with noise from its orthogonal counterpart in a
frequency-dependent manner~\cite{Kwee14a}.

Cavity birefringence~\cite{BoonEngering1997,Asenbaum11} was
investigated as a possible source of frequency-dependent
loss. However, studies verified that the elliptical birefringence in
our system, which unlike linear birefringence cannot easily be
countered, was negligible. This effect should be revisited in the
context of any future systems.

Mitigating the above-mentioned technical noise effects, rather than
concentrating on generating stronger squeezing at the source, is
currently the most profitable route toward improved performance.

\subsection{Scaling for Gravitational Wave Detectors}

While this demonstration of frequency dependent squeezing has brought
the squeezed quadrature rotation frequency 4~orders of magnitude
closer to that required by \gw\ detectors, it is still a factor of
$\sim$20 away from the \SI{50}{Hz} target of Advanced LIGO.  It is
foreseen, however, that the initial implementation of frequency
dependent squeezing in Advanced LIGO will involve a filter cavity
\SI{16}{m} in length, or a factor of 8 longer than the cavity used in
this demonstration~\cite{Eva13a}.  Furthermore, detailed measurements
of losses in long-storage time cavities~\cite{Isogai13}, and
calculations of the impact of these losses on the performance of
frequency dependent squeezing~\cite{Kwee14a}, indicate that a
\SI{16}{m} cavity with finesse roughly 3 times that of the one used in
the present work will be sufficient for Advanced LIGO.  Such a filter
cavity would have a storage time of \SI{2.5}{ms}, and
$\tau_\text{decoherence} \simeq \SI{0.7}{ms}$, which is sufficient to
maintain a modest level of squeezing below the rotation
frequency~\cite{Eva13a,Kwee14a}.

Based on the results presented here, previous experimental work \cite{Bar13a,Isogai13,Dooley2015}
 and extensive theoretical studies \cite{Eva13a,Kwee14a,Oelker2014},
 the authors and other members of the LIGO Laboratory have begun the process of
 designing and building a full-scale prototype frequency dependent squeezed light source for Advanced LIGO.

\section{Conclusions}
\label{sec:Conclusions}


The principal goal of this endeavor was to demonstrate
frequency-dependent quadrature rotation in a band relevant to \gw\
detectors, informing the design of all future squeezed light sources
in the field. A frequency-independent squeezed vacuum source is only
able to reduce noise in the band in which its (fixed) low-noise
quadrature is well-aligned to the interferometer signal field.  In
this case, the observed noise reduction would be approximated by a
single one of the curves shown in \Fref{results}. For example,
squeezing the quadrature phase, as previously demonstrated in LIGO and
GEO600, would reduce noise at high frequency and increase noise at low
frequency, as described by the blue, $\phi = 90^\circ$,
curve. The equivalent frequency-dependent source offers equal
performance at high frequency but a relative improvement of nearly a
factor of 10 in strain amplitude at low frequency.

Applied to a \gw\ detector whose optomechanical response
conforms with our filter cavity rotation, and assuming the same
$\sim$\SI{30}{\percent} total losses, the overall noise reduction
available from this frequency-dependent squeezed vacuum source is
given by the solid black curve in \Fref{results}. This trace
represents the lower envelope of the recorded data and the infinite
family of curves at intermediate quadratures.

All present ideas for extending the reach of ground-based \gw\
observation rely on audio-band frequency-dependent
squeezing~\cite{Rana:RMP, Lungo14, ET}. We have demonstrated
quadrature rotation of squeezed vacuum at audio frequencies, bringing
frequency dependent squeezing 4~orders of magnitude closer to the
required frequency regime.  Moreover, our measurement uses a
relatively short filter cavity and thus leaves a clear path toward
scaling to longer storage times.

Extrapolating our results to the case of Advanced LIGO, assuming
parameters within reach of current technology, we find that the
reduction of quantum noise with frequency-dependent squeezing
increases the volume of the detectable universe by about a factor of
two~\cite{Miller2015a}. Larger gains, up to nearly a factor of 10 in
volume, are achievable when frequency dependent squeezing is combined
with other improvements~\cite{Miller2015a}.



\vspace{10pt}
\begin{acknowledgments}
The authors wish to acknowledge the contribution made by Patrick
  Kwee to the design and construction of the filter cavity apparatus
  and to thank Myron MacInnis for providing technical support.  The
  initial design of the OPO used in this experiment was provided by
  the group led by David McClelland at The Australian National
  University.  Valuable input was received from the LIGO Scientific
  Collaboration's Advanced Interferometer Configurations and Quantum
  Noise working groups, in particular from Daniel Sigg, Sheila Dwyer
  and Peter Fritschel.  We also thank Vladan Vuletic for fruitful
  conversations, and Jan Harms for carefully reading the manuscript.
 The authors gratefully acknowledge the support of
  the National Science Foundation and the LIGO Laboratory, operating
  under cooperative Agreement No. PHY-0757058. This research was
  supported in part by an award to E.O. from the Department of Energy
  Office of Science Graduate Fellowship Program under DOE contract
  number DE-AC05-06OR23100.
 \end{acknowledgments}



\bibliography{papers}

\end{document}